\documentclass[%
 reprint,
showpacs,
 amsmath,amssymb,
 aps,
 pra,
longbibliography,
floatfix
]{revtex4-1}

\usepackage[utf8]{inputenc}	%Tillader danske tegn
\usepackage[T1]{fontenc}
\usepackage{graphicx}		%Tillader indsættelse af billeder
\usepackage{color}
\usepackage[english]{babel}
\usepackage{hyperref}
\usepackage{url}
\usepackage{todonotes}
\usepackage{lipsum}
\usepackage{bm}
\usepackage{microtype}

\usepackage{braket}

\usepackage[position=top,caption=false]{subfig}

\let\originaleqref\eqref
\renewcommand{\eqref}{Eq.~\originaleqref}

\newcommand{\fref}[1]{Fig.~\ref{#1}}

\newcommand{\e}[1]{\mathrm{e}^{#1}}

\newcommand{\Tr}[1]{\mathrm{Tr}\left(#1\right)}

\newcommand{\C}[0]{\hat{c}}
\newcommand{\Cd}[0]{\hat{c}^\dagger}

\renewcommand{\L}[0]{\hat{L}}
\newcommand{\Ld}[0]{\hat{L}^\dagger}

\renewcommand{\sp}[0]{\hat{\sigma}_\mathrm{+}}
\newcommand{\sm}[0]{\hat{\sigma}_\mathrm{-}}

\newcommand{\tp}{t^\prime}

\renewcommand{\b}{\hat{b}}
\renewcommand{\c}{\hat{c}}
\renewcommand{\a}{\hat{a}}
\newcommand{\ad}{\hat{a}^\dagger}
\newcommand{\bd}{\hat{b}^\dagger}
\renewcommand{\H}{\hat{H}}

\graphicspath{
{..figures/}
{/home/alexander/Dropbox/PhD/generalizedInputOutput/figures/}
}

\begin{document}
\title{Quantum interactions with pulses of radiation}
\author{Alexander Holm Kiilerich}
\email{kiilerich@phys.au.dk}
\author{Klaus Mølmer}
\email{moelmer@phys.au.dk}
\date{\today}
\affiliation{Department of Physics and Astronomy, Aarhus University, Ny Munkegade 120, DK 8000 Aarhus C. Denmark}
\date{\today}

\bigskip

\begin{abstract}
This article presents a general master equation formalism for the interaction between travelling pulses of quantum radiation and localized quantum systems.
Traveling fields populate a continuum of free space radiation modes and the Jaynes-Cummings model, valid for a discrete eigenmode of a cavity, does not apply.
We develop a complete input-output theory to describe the driving of quantum systems by arbitrary incident pulses of radiation and the quantum state of the field emitted into any desired outgoing temporal mode. 
Our theory is applicable to the transformation and interaction of pulses of radiation by their coupling to a wide class of material quantum systems. We discuss the most essential differences between quantum interactions with pulses and with discrete radiative eigenmodes and present examples relevant to quantum information protocols with optical, microwave and acoustic waves.    
\end{abstract}

\maketitle
\noindent

\section{Introduction}
Many quantum technologies rely on the preparation and interaction of pulses of radiation with matter. In particular, in the field of quantum information processing and communication \cite{kimble2008quantum,o2009photonic}, quantum state transfer between stationary and travelling physical components are gaining importance, see, e.g., \cite{parkins1999quantum,cirac1997quantum,matsukevich2004quantum,zhang2003quantum,PhysRevLett.118.133601,kimble2008quantum,stute2013quantum,vermersch2017quantum,vermersch2017quantum}. While a host of experimental and theoretical results on the basic quantum interactions between light and matter is now textbook material, researchers have only recently undertaken efforts to properly describe the interaction of quantum systems with propagating wave packets of light and other forms of radiation. For a recent review, see \cite{fischer2018scattering}.

\begin{figure}[h!]
\centering
\includegraphics[trim=0 0 0 0,width=0.95\columnwidth]{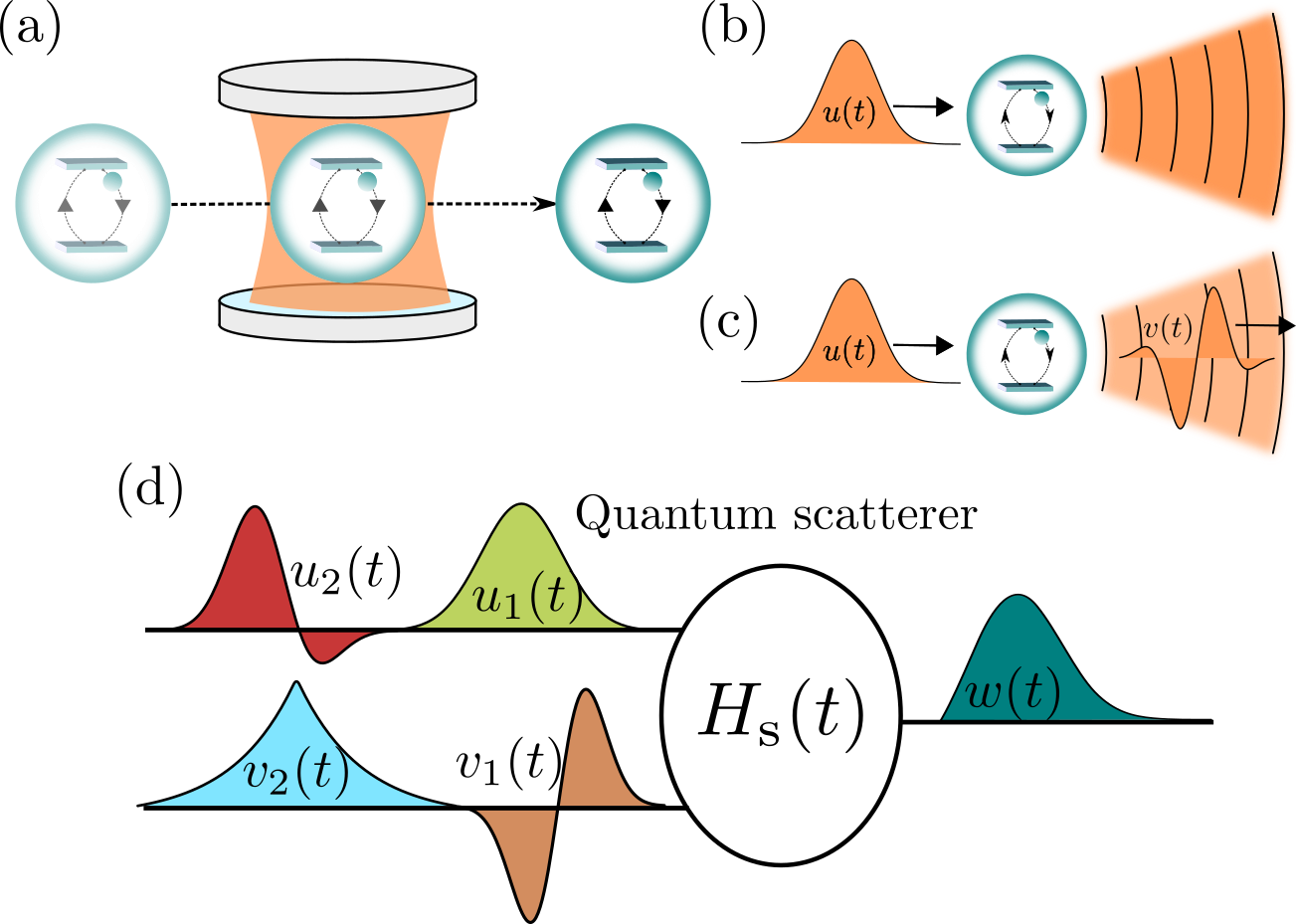}
\caption{Interaction, a) between a flying two-level system and the quantum field in a cavity,  b) between a stationary two-level atom and an incident pulse, resulting in atomic excitation and scattering, c) between a two-level atom and an incident pulse, with an appreciable fraction of the outgoing energy occupying a single output pulse, d) setup where multiple input pulses excite a quantum system and cause reflection and transmission into multiple output pulses.}
\label{fig:setup}
\end{figure}

Standard quantum optics textbooks discuss non-classical properties of light through the introduction of quantum states such as Fock (number), coherent and squeezed states, and introduce quantized light-matter interactions by the seminal Jaynes-Cummings model, 
\begin{equation} \label{eq:HJC}
\hat{H}_{\textit{JC}} =  g(\hat{a}\sp + \hat{a}^{\dagger}\sm),
\end{equation}
where $\hat{a}^{(\dagger)}$ and $\hat{\sigma}_{(+)-}$ are the (creation)annihilation operators of the photon field and the excitation of a two level quantum system and $g$ is the coupling strength. Figure~\ref{fig:setup}(a) illustrates the realization of this model by a two level system passing through the field confined in a cavity (the coupling $g(t)$ is then time dependent as the atom traverses regions with different strengths of the electromagnetic field mode).

Cavity systems and their equivalents in circuit QED support discrete modes which may justify the restriction of quantum interactions to only a single resonant mode as in Eq.~(\ref{eq:HJC}). But travelling fields explore a continuum of modes which simultaneously incorporate Maxwell's equations of wave propagation and the (second quantization) concept of creation and annihilation operators. In linear media, travelling Maxwell wave packets merely propagate their quantum state contents,
and one might expect that light in an incoming wave packet $u(t)$, see Fig.~\ref{fig:setup}(b), would interact with a two-level system in the same way as a moving atom interacts with a stationary eigenmode of light. However, when the light interacts with a non-linear medium such as a two-level system, the photon number contents and the wave packet shape may change in a correlated manner and thus explore the full multimode character of the quantized field. This does not happen in the cavity if a large frequency gap suppresses coupling to other eigenmodes.

A quantum system, such as a two-level atom coupled to a continuum of radiation modes in the vacuum state, can be effectively described as an open quantum system and the corresponding reduced master equation can be solved for the system density matrix. That equation permits inclusion of driving by a classical pulse as a time dependent term in the system Hamiltonian. The field emitted by the system can be characterized by the time dependent amplitude and intensity whose mean values are governed by the atomic coherence and excited state population, see Fig.~\ref{fig:setup}(b). This method, however, does not permit description of the excitation of the system by a light pulse prepared in a non-classical state. Moreover, the mean field and intensity neither provide the full Schr\"odinger picture quantum state of the emitted field nor the quantum state contents of any subset of propagating field modes. 

It is possible to describe quantum interactions and propagation in quantum media and explore the full quantum state evolution of pulses with up to two excitations \cite{Motzoi_2018,PhysRevA.92.033803, PhysRevA.82.063821,witthaut2010photon,witthaut2012photon,nisbet2013photonic,PhysRevA.82.033804,PhysRevA.82.033804,PhysRevLett.114.173601,bock2018high,
PhysRevA.92.053834,Caneva_2015,PhysRevA.99.033835}. 
The present article reviews and expands an alternative theoretical treatment, introduced in a recent Letter \cite{PhysRevLett.123.123604}. 
This theory provides the full quantum state only for individually specified modes [see Figs.~\ref{fig:setup}c)~and~d)] but is not restricted by the number of excitations.   
At the same time, it has a more straightforward interpretation and is easier to implement than the recent Fock state master equation approach of Baragiola \textit{et. al.} \cite{PhysRevA.86.013811}, and alternative formulations \cite{gheri1998photon,gough2012single,PhysRevA.86.043819} derived from Itô calculus. Furthermore, unlike these approaches, we can provide the quantum state of both input and output field modes. While our focus is on few mode quantum states, mean values and correlation functions of the radiation components outside these modes, indicated by the shaded wave fronts in Fig.~\ref{fig:setup}(c), are also directly available in our formalism. 

The aim of the present manuscript is to develop the theory and highlight some of the main physical differences between quantum interactions with a stationary mode and a travelling pulse of quantum radiation. In Sec. II, we present the basic theory, and we  offer examples of its application to describe decoherence of a quantum pulse and production of pulses of non-classical radiation. In Sec. III, we present a generalization to multimode pulses, illustrated by emission of quantum pulses that are entangled with the final state of the emitter. In Sec. IV, we apply our theory to a paradigmatic photon blockade proposal by a cavity with a single atom and we identify a fundamental time-bandwidth restriction on non-linear quantum optical schemes with pulses of radiation. In Sec. V, we describe how to model pulses propagating through a waveguide in a thermally excited state. In Sec. VI, we conclude and discuss future prospects of the method.  

\section{Theory}\label{sec:core}
Consider a quantum system described, in the Born-Markov approximation, by a Hamiltonian $\H_\mathrm{s}$ and a set of $n$ dissipation operators $\{\L_i\}_{i = 1}^n$ such that the evolution of its quantum state $\rho_{\mathrm{s}}$ is governed by the master equation ($\hbar=1$)
\begin{align}\label{eq:me}
\frac{d\rho}{dt} = \frac{1}{i}\left[\H,\rho\right] + \sum_{i=1}^n \mathcal{D}[\hat{L_i}]\rho,
\end{align}
with $\H = \H_\mathrm{s}$ and $\mathcal{D}[\hat{L_i}]\rho = \L_i\rho\Ld_i-\frac{1}{2}\left(\Ld_i \L_i\rho+\rho\Ld_i \L_i\right)$.
In addition, the system interacts with the quantized field via $V = i\sqrt{\gamma}(\bd_{\mathrm{in}}\c-\b_{\mathrm{in}}\Cd)$, where $\c$ annihilates an excitation in the system and $\b_{\mathrm{in}}(t)$ is the annihilation operator of the multimode bosonic input field which obeys the commutation relations $[\b_{\mathrm{in}}(t),\bd_{\mathrm{in}}(t')]=\delta(t-t')$, such that $\langle \bd_{\mathrm{in}}(t)\b_{\mathrm{in}}(t)\rangle$ is the rate of photons incident on the local quantum system at time $t$. 

The conventional input-output theory of quantum optics \cite{Gardiner1985, GardinerBook} provides an operator expression for the output field 
\begin{align}\label{eq:inout}
\b_{\mathrm{out}}(t) = \b_{\mathrm{in}}(t) + \sqrt{\gamma} \c,
\end{align}
connecting the asymptotic incoming and outgoing field component that spatially overlap with the scatterer system at time $t$. If the system Hamiltonian $\H_\mathrm{s}$ is at most quadratic and the damping is linear in bosonic annihilation and creation operators, the Heisenberg equations of motion for the system operator $\c$ can be solved and $\b_{\mathrm{out}}(t)$ expressed in terms of the input fields. This is, however, not the case for scattering on few-level, anharmonic or nonlinear systems.

Our theory combines the input-output theory with the concept of cascaded quantum systems \cite{PhysRevLett.70.2269,Carmichael1993}, which describes how the output from one system can serve as an input to another system while formally eliminating the propagating quantum field modes from the theory.
To describe an incident wavepacket $u(t)$, we introduce a theoretical model with a leaking cavity which emits the corresponding wave packet containing the initial quantum state of the cavity mode. Similarly, the quantum state of any specific outgoing wave packet can be modelled as the state transferred into a single cavity mode, forming also a discrete component in our theory~\cite{PhysRevLett.123.123604}. By this method we obtain an effective master equations for the density matrix of the discrete quantum system and the input and output pulses described by two (pseudo-cavity) modes.

We restrict the formal theory to one dimensional wave propagation, i.e., we assume a waveguide or a collimated beam, with only a single transverse mode. We also assume a chiral coupling of the components: the radiation propagates towards and away from the scatterer along distinct input and output directions [along the arrows in Fig.~\ref{fig:setup}(b-c)]. Reflection and transmission may, however, be treated as separate output channels [Fig.~\ref{fig:setup}(d)].

\subsection{Driving with a quantum pulse}
If a single mode cavity is coupled to an input field with amplitude $g(t)$, the quantum Langevin equation
for the field operator $\a$ reads \cite{GardinerBook}
\begin{align}\label{eq:QLE}
\dot{\a} = - \frac{|g(t)|^2}{2}\a - g(t) \b_{\mathrm{in}}(t),
\end{align}
where we assume a rotating frame around the carrier frequency of the field mode. Note that if $g(t)$ varies slowly compared to the spectral range of the continuum field, the Born-Markov approximation gives the time dependent cavity decay rate $|g(t)|^2$.
The general solution for the intra-cavity field reads
\begin{align}\label{eq:QLEsol}
\begin{split}
\a(t) &= \e{-\frac{1}{2}\int_0^{t} dt' \, |g(t')|^2}\a(0)
\\
&-\int_0^{t} dt' \, g(t')\e{-\frac{1}{2}\int_{t'}^{t} dt''\, |g(t'')|^2}\b_{\mathrm{in}}(t').
\end{split}
\end{align}
The input-output relation (\ref{eq:inout}) yields $\bd_{\mathrm{out}}(t)=
\bd_{\mathrm{in}}(t)+g^*(t)\ad(t)$, and we can define the creation operator 
\begin{align}\label{eq:bu}
b_u^\dagger = \int dt\, u(t) \bd_{\mathrm{out}}(t)
\end{align}
for the temporal output mode of the cavity with envelope
$u(t) = g_u^*(t)\e{-\frac{1}{2}\int_0^{t} dt' \,  |g_u(t')|^2}$, normalized as $\int dt\, |u(t)|^2 = 1$. Upon inverting this expression, one finds \cite{gough2015generating} the  time-dependent coupling
\begin{align}\label{eq:gu}
g_u(t)  = \frac{u^*(t)}{\sqrt{1-\int_0^t dt'\, |u(t')|^2}},
\end{align}
required for the cavity field to be emitted in the wave packet $u(t)$.

For a quantum system, it is equivalent to be driven by a travelling pulse and by the output field of a cavity, and according to the theory of cascaded quantum systems, the joint state $\rho_{\mathrm{us}}$ of the cavity with field annihilation operator $\hat{a}_u$ and the quantum system is described by a master equation of Lindblad form~(\ref{eq:me})
%\begin{align}\label{eq:me}
%\frac{d\rho}{dt} = -i[\H,\rho]+ \sum_{i=0}\left( \L_i\rho \L_i^\dagger-\frac{1}{2}\left\{\L_i^\dagger \L_i, \rho\right\}\right),
%\end{align}
%where $\{\cdot,\cdot\}$ denotes the anti-commutator, 
with a Hamiltonian given by
\begin{align}\label{eq:Hus}
\begin{split}
&\H_\mathrm{us}(t) = \H_{\mathrm{s}}(t)+\frac{i\sqrt{\gamma}}{2}\big(g_u^*(t)\a_u^\dagger\c - g_u(t)\a_u\c^\dagger \big).
\end{split}
\end{align}
In addition to the Lindblad terms acting only on the quantum system in Eq.~(\ref{eq:me}), the system and the input cavity mode are subject to a time dependent Lindblad term $\mathcal{D}[\hat{L}^{(\mathrm{us})}_0(t)]$ with operator
\begin{align}\label{eq:Lus}
\hat{L}^{(\mathrm{us})}_0(t) = g_u(t)\a_u + \sqrt{\gamma}\c.
\end{align}
Equation~(\ref{eq:Lus}) is of the same form as the input-output relation~(\ref{eq:inout}), and indeed represents the output field from the quantum system, composed of interfering contributions from the input field and the emission by the system itself. 

Combining the Hamiltonian and Lindblad terms in the master equation, we obtain the master equation,
\begin{align}\label{eq:me_rewritten}
\begin{split}
\dot{\rho}_\mathrm{us} &=-i\left[\H_\mathrm{s},\rho\right] + \sum_{i=1}^n \mathcal{D}[\hat{L_i}]\rho
\\
&+ \sqrt{\gamma}\big[g_u(t)(\a_u\rho_\mathrm{us}\Cd -\a_u \Cd \rho_\mathrm{us}) 
\\
&+ g^*_u(t)(\c\rho_\mathrm{us}\ad_u-\rho_\mathrm{us} \ad_u\c)\big]
\\
&+ \mathcal{D}[\sqrt{\gamma}\c]\rho_\mathrm{us} + \mathcal{D}[g_u(t)\a_u]\rho_\mathrm{us}.
\end{split}
\end{align}
This equation deals explicitly with a density matrix which spans the tensor product Hilbert space of the quantum system and the input pulse cavity mode, and has the same dimension and numerical complexity as the Fock space master equation by Baragiola \textit{et. al.} \cite{PhysRevA.86.013811}.

Note that in \eqref{eq:me_rewritten}, the Hamiltonian and the $\mathcal{D}[\hat{L}^{(\mathrm{us})}_0(t)]$ damping terms conspire such that $\rho_\mathrm{us}$ is only operated upon from the left(right) by $\a_u(\ad_u)$: quanta are only annihilated from the incoming pulse and never created in it, signifying the cascaded nature of the scattering process. 

The formal structure of Eq.~(\ref{eq:me_rewritten}) implies that if the input mode is initially prepared in a coherent state $\ket{\alpha}$ with $\a_u\ket{\alpha} = \alpha\ket{\alpha}$, it remains in a coherent state with a time dependent amplitude $\alpha(t)$, damped according to $\dot{\alpha}(t) = -  \frac{|g_u(t)|^2}{2}\alpha(t)$. The quantum state then factorizes and we get a reduced master equation for the discrete system with the time dependent Hamiltonian
\begin{align}
\hat{H} = \H_\mathrm{s} + i\sqrt{\gamma}\left[u(t)\alpha^*(0)\c-u^*(t)\alpha(0) \Cd\right],
\end{align}
describing the interaction with a classical, time dependent field, the dissipation terms of \eqref{eq:me}, and spontaneous decay into the propagating field governed by a single Lindblad operator
\begin{align}\label{eq:Ls}
\hat{L}^{(\mathrm{s})}_0(t) = \sqrt{\gamma}\c,
\end{align}
(see also Ref.~\cite{PhysRevX.7.041010}). This dynamics, in fact, becomes even simpler than the interaction with a single mode cavity field in a coherent state, for which the Jaynes-Cummings model (\ref{eq:HJC}),  yields complex dynamics with collapses and revivals of Rabi oscillation due to the different coupling amplitudes $g\sqrt{n}$ of the transitions involving different photon number components, $|g,n\rangle \leftrightarrow |e,n-1\rangle$. For input quantum states other than coherent states, however, we have recourse to Eq.~(\ref{eq:me_rewritten}) to make predictions for the time evolution of the interacting systems.

\subsection{Output field mean values and correlation functions}
\label{sec:g1}
Our formalism determines the state of a quantum system subject to an input pulse, but the field component of $\rho_{\mathrm{us}}$ eventually converges to the vacuum state and does not describe the state of the output field. 
We can obtain mean values and higher moments of the scattered field via the input-output relation~(\ref{eq:inout}). In particular, the time dependent intensity is given by
\begin{align}\label{eq:intensity}
I_{\mathrm{out}}(t) = \braket{[\hat{L}^{(\mathrm{us})}_0(t)]^\dagger \hat{L}^{(\mathrm{us})}_0(t)},    
\end{align}
while the autocorrelation function of the output field
$
g^{(1)}(t,t') = \langle (\L^{(\mathrm{us})}_0(t))^\dagger\hat{L}^{(\mathrm{us})}_0(t')\rangle
$
and, thus, its spectrum are given by the quantum regression theorem \cite{breuer2002theory,GardinerBook} as
\begin{align}\label{eq:g1}
g^{(1)}(t,t') = \mathrm{Tr}\left\{[\L^{(\mathrm{us})}_0(t)]^\dagger\Lambda(t,t')\left[\L^{(\mathrm{us})}_0(t)\Lambda(t',0)\rho_{\mathrm{us}}(0)\right]\right\}.  
\end{align}
Here $\rho_{\mathrm{us}}(0)$ is the joint state of the incoming mode and the quantum system and $\{\Lambda(t,t')\}_{t\geq 0}$ represents the linear time evolution map of the master equation~(\ref{eq:me_rewritten}). 
Finally, we note that the eigenmode decomposition, 
\begin{align}\label{eq:g1decom}
g^{(1)}(t,t') = \sum_i n_i v_i^*(t)v_i(t')
\end{align}
of the autocorrelation function determines the most occupied set of orthogonal modes $v_i(t)$ in the output field with $n_i$ quanta of excitation.

%%%%%%%%%%%%%%%% CAVITY WITH PHASE NOISE EXAMPLE %%%%%%%%%%%%%%%%

\begin{figure}
\centering
\includegraphics[trim=0 0 0 0,width=0.95\columnwidth]{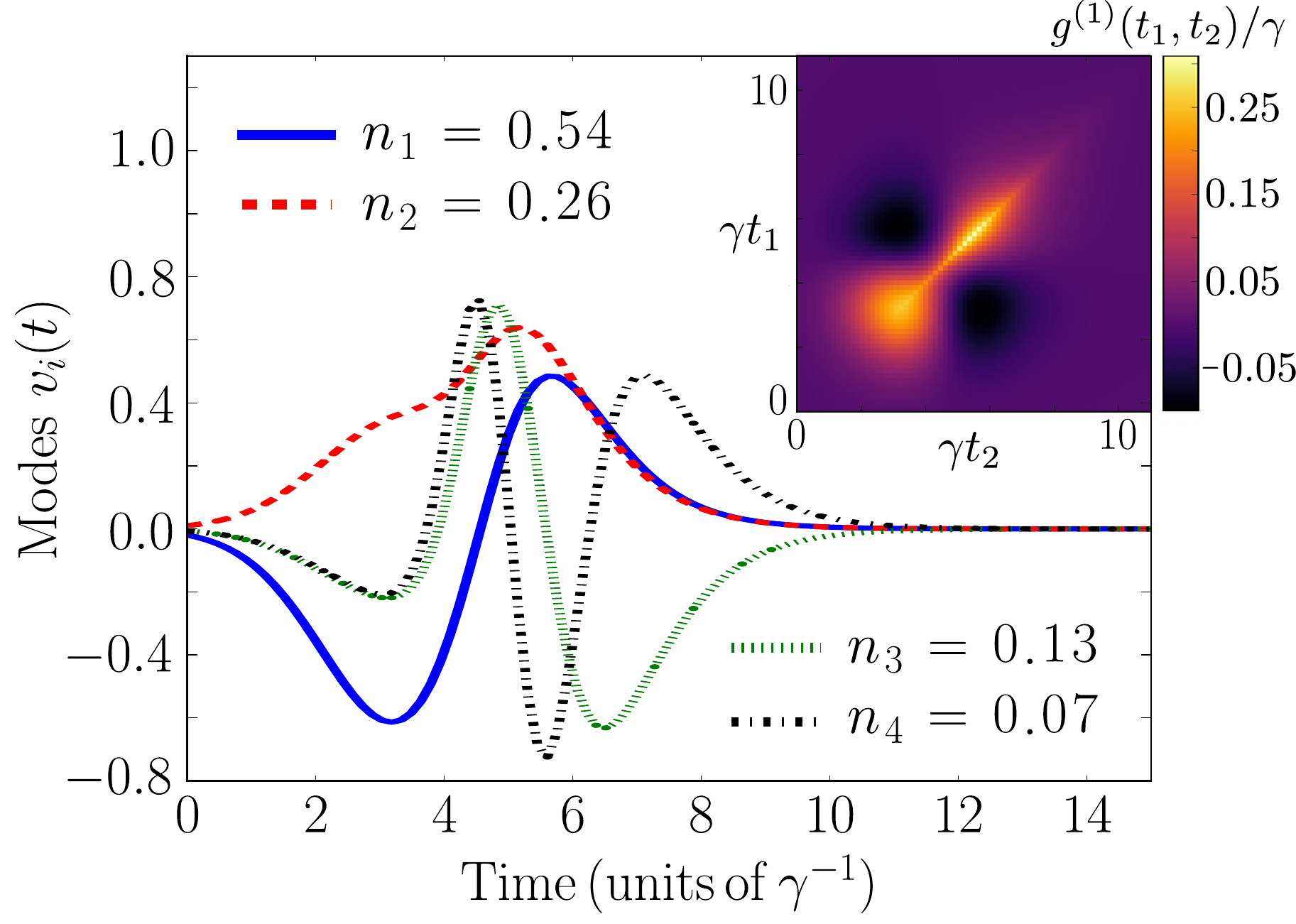}
\caption{
Scattering on an empty cavity with phase noise.
The four dominating orthogonal modes $v_1(t)$, $v_2(t)$, $v_3(t)$ and $v_4(t)$ in the output field found from the autocorrelation function $g^{(1)}(t_1,t_2)$ for the field emitted by the cavity (shown in the inset).
The respective fractions, $n_1$, $n_2$,$n_3$ and $n_4$, of the input photon number are given in the legend.
Results are shown for a Gaussian input mode~(\ref{eq:uGauss}) of duration $\tau = \gamma^{-1}$ arriving at $t_p = 4\gamma^{-1}$ and a cavity phase fluctuation rate of $\gamma_p = 1.5\gamma$.
% Figure at 
%/home/alexander/Dropbox/PhD/generalizedInputOutput/python/modeSelection/examples/figures
}
\label{fig:phaseNoise}
\end{figure}

In the simple example of scattering of a pulse $u(t)$
on an empty cavity with resonance frequency $\omega_c$ (the local quantum system), the quantum state of the pulse is unchanged but the output populates a modified mode related to the input pulse by the frequency domain expressions, $v(\omega) = [i(\omega-\omega_c)+\gamma/2]/[i(\omega-\omega_c)-\gamma/2]u(\omega)$.  
If, however, the cavity experiences phase noise, as described by an additional Lindblad term $\hat{L}_1 = \sqrt{\gamma_p}\Cd \C$ in the master equation~({\ref{eq:me}}) for the cavity mode, the output field occupies several orthogonal modes. We have calculated $g^{(1)}(t_1,t_2)$ (see inset in \fref{fig:phaseNoise}) and identified the four output modes with the largest populations in \fref{fig:phaseNoise} for$\gamma_p=1.5\gamma$ and an input pulse of Gaussian shape,
\begin{align}\label{eq:uGauss}
u_{\mathrm{Gauss}}(t) = \frac{1}{\sqrt{\tau}\pi^{1/4}}\e{\frac{-(t-t_p)^2}{2\tau^2}}    
\end{align}
with $\tau=\gamma^{-1}$.
The modes are orthogonal and have completely different characteristics and the output field is distributed over many more modes, cf., the populations in the first nine modes:  $
\{n_i\}_{i=1}^9 =
\{
0.54,
0.26,
0.13,
0.072,
0.047,
0.033,
0.025,
0.019,
0.015
\}.
$
%%%%%%%%%%%%%%%%%%%%%%%%%%%%%%%%%
\subsection{Scattering into a quantum pulse}
\label{sec:scatToQPulse}\label{sec:vstate}

For a growing number of applications, it is pertinent to obtain the quantum state rather than mean values of the scattered field and mean occupation of the dominant eigenmodes.
Often we are interested in the quantum state of a single or a few dominant output pulse mode functions, which may, for example, be chosen among the most populated orthogonal modes, identified by Eq.~(\ref{eq:g1decom}). Our cascaded system master equation may readily provide full information about the quantum state contents of any outgoing wave packet mode (or few modes) while treating the emission into other modes as losses.  

To obtain a full quantum state description of a chosen output mode $v(t)$, we introduce another downstream virtual cavity with a time dependent coupling $g_v(t)$. Assuming a complete asymptotic decay of the initial amplitude in this cavity, i.e., that the first term in \eqref{eq:QLEsol} vanishes, the integral over the input field in the second term for $t\rightarrow \infty$ has the temporal weight factor $v(t) = -g_v^*(t)\e{-\frac{1}{2}\int_{t}^{\infty} dt'\, |g_v(t')|^2})$. To fully capture the desired pulse, we require \cite{7798639}
\begin{align}\label{eq:gv}
g_v(t) = -\frac{v^*(t)}{\sqrt{\int_0^t dt'\, |v(t')|^2}}.
\end{align}
The virtual output cavity is cascaded after the localized quantum system such that the full quantum state $\rho\equiv\rho_{\mathrm{usv}}$ now represents three components; the cavity releasing the incoming pulse, the quantum system exposed to the field, and the cavity capturing the outgoing pulse.

If we denote the annihilation operator of the output mode $\a_v$, the master equation~(\ref{eq:me}) would apply to the full system with the Hamiltonian 
\begin{align}\label{eq:H}
\begin{split}
&\H(t) = \H_{\mathrm{s}}(t)+\frac{i}{2}\big(\sqrt{\gamma}g_u^*(t)\a_u^\dagger\c
\\
&+ \sqrt{\gamma^*} g_v(t)\c^\dagger\a_v  + g_u^*(t) g_v(t)\a_u^\dagger\a_v - \mathrm{h.c.}\big),
\end{split}
\end{align}
The system damping terms in~(\ref{eq:me}) are now supplemented by $\mathcal{D}[\hat{L}_0(t)]$ with
\begin{align}\label{eq:Luv}
\hat{L}_0(t) = \sqrt{\gamma}\c + g_u(t)\a_u+g_v(t)\a_v. 
\end{align}

If the scattered field is fully accommodated by the mode $v(t)$, the cascaded network evolves along a dark state of the dissipator $\hat{L}_0$, while mismatch of the mode $v(t)$ with the output field results in loss with a rate $I_\mathrm{out}(t) = \braket{\hat{L}^\dagger_0(t) \hat{L}_0(t)}$. The emission of quanta into other modes from time $t_1$ to $t_2$ may thus be found by evaluating $\int_{t_1}^{t_2} dt\, I_\mathrm{out}(t)$.

%%%%%%%%%%%%%%%%% NAKAMURAS CAT %%%%%%%%%%%%%%%%%%%%%%%%%%%%%%%%
\subsection{Production and release of a non-classical  pulse of radiation}\label{sec:nakamurastate}

Interesting quantum states of light can be produced by the classical driving of non-linear quantum systems, i.e., without the need of an incident quantum pulse. In many cases, however, these states have only been characterized by low order correlation functions, or rather complicated reconstruction of the quantum state has been accomplished via a hierarchy of operator moments~\cite{PhysRevA.99.023838} or simulated tomography~\cite{PhysRevA.100.063808}.  
To obtain the quantum states of the output field in such situations with our more straightforward method, we omit the degrees of freedom of the input cavity in Eqs.~(\ref{eq:H})~and~(\ref{eq:Luv}), and study only the emitter quantum system and the cavity extracting the output mode of interest. 

As an example, we consider the on-demand generation of a traveling Schrödinger cat state by a Kerr-nonlinear parametric oscillator (KPO) driven by a classical pump field as described by the Hamiltonian~\cite{PhysRevA.99.023838},
\begin{align}
H_{\mathrm{s}}(t) = \frac{p(t)}{2}\left[(\ad)^2+\a^2\right]-\frac{K}{2}(\ad)^2 a^2+\Delta \ad \a
\end{align}
Here $\Delta$ is the pump detuning which we set to zero, $p(t)$ is the time dependent pump amplitude, and $K$ is the magnitude of the Kerr coefficient which is assumed negative in Ref.~\cite{PhysRevA.99.023838}.

If the KPO is a closed system, a cat state of the cavity field, $\ket{\text{cat}} = \frac{1}{\sqrt{2}}\left(\ket{\alpha_0}+\ket{-\alpha_0}\right)$ is adiabatically generated from the vacuum state by gradually increasing $p(t)$ from zero to $p_0 = K\alpha_0$.
When the KPO is coupled to the output field, however, the field leaks out during the generation of the state and it is not clear if a cat state will ultimately populate a single wave packet mode. Reference~\cite{PhysRevA.99.023838} suggests that this will be the case if the cat is prepared much faster than the cavity decay (assuming $K$ is much larger than the cavity decay rate $\gamma$) and the pump is gradually switched off as $p(t) = p_0\e{-\gamma(t-t_0)}$ after the time $t_0$ where the KPO cavity mode has ideally reached a cat state. By a multi-time correlation function analysis, Ref.~\cite{PhysRevA.99.023838}, indeed, demonstrates that a travelling pulse cat state is prepared under these conditions. 

The production of a travelling cat state by the KPO provides an ideal test for our theory. Following Ref.~\cite{PhysRevA.99.023838}, we set $K = 5\gamma$ and let the pump originate from a fourth-order low pass filter (LPF) with vanishing input for $t\leq 0$ and $p_{\mathrm{in}}(t) = K A_p\e{-\gamma t}$ for $t > 0$.
(The output of a LPF with bandwidth $B$ is given by
$p_{\mathrm{out}}(t) = \int_0^t dt'\, B\e{-B(t-t')}p_{\mathrm{in}}(t')$, and an $n$th-order filter is defined by feeding the output into a new filter $n-1$ times. We set $B = 2.5\gamma$ and $A_p\simeq 4.45$  to fix the photon number $|\alpha|^2$ to a value around $4$ in the cat state produced.) 
%The final time is set such that the KPO is basically empty at the end of the experiment (population less than $10^{-3}$). 
%Notice that $\dot{p} = B[p(t)-p_{\mathrm{in}}(t)]$ which is probably an easier way to construct $p(t)$. Parameters are given in a table in the paper.

Following our description above, we first solve the master equation of the driven and damped KPO in order to determine the most populated output mode $v(t)$ from the cavity field autocorrelation function $g^{(1)}(t,t^\prime)$, 
%where $I_2(t) = \gamma{\ad a}$ is the time dependent intensity of the emission from the KPO which is initially evaluated by solving the master equation~(\ref{eq:mes}) for the KPO alone. 
 
For the parameters used here, the dominating mode $v(t)$ acquires $4.03$ photons while less than $0.05$ photons appear in other propagating modes.

\begin{figure}
\centering
\includegraphics[trim=0 0 0 0,width=0.95\columnwidth]{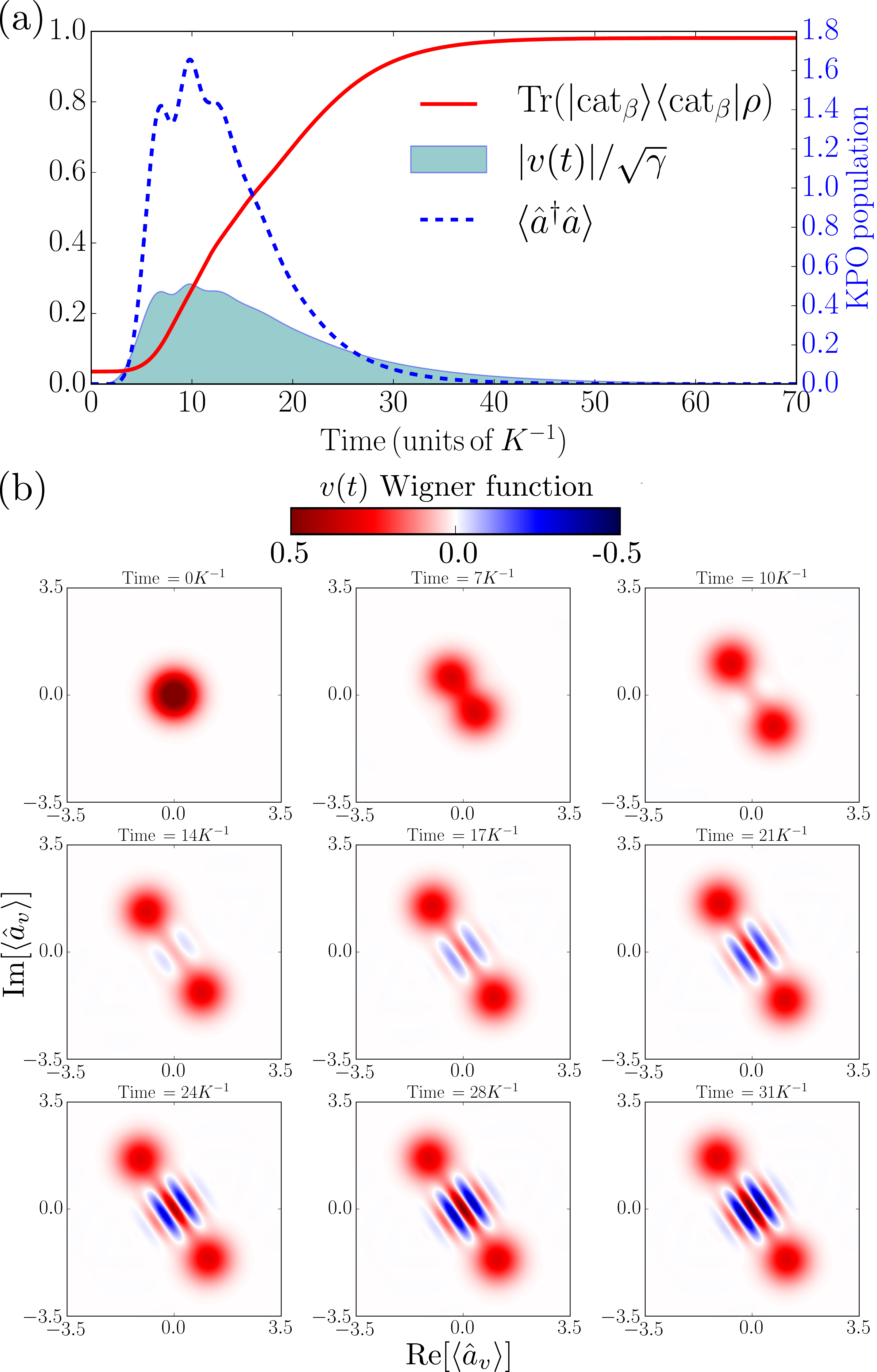}
\caption{Travelling Schrödinger cat state generated by a classically driven KPO.
(a) Red solid curve (left axis): Fidelity of the cat state~(\ref{eq:betaCat}) in the output mode $v(t)$. Blue dotted curve (right axis): Excitation in the KPO during the operation. Shaded area: the shape of the most occupied output mode $v(t)$. 
(b) Wigner function of the field captured by the output cavity (ultimately the contents of the output mode $v(t)$) at different times, annotated above each panel. 
Results are shown for $K=5\gamma$, $B = 2.5 \gamma$ and $A_p= 4.45$.}
\label{fig:catA}
\end{figure}

After identifying the most populated mode, we solve the cascaded system master equation for the KPO and the corresponding $\hat{a}_v$-cavity. The results are shown in  \fref{fig:catA}, where panel (a) shows  
the excitation in the KPO cavity, the shape of the most occupied output mode $v(t)$, and the cat state fidelity as functions of time. The fidelity of the cat state (red line) is defined as  
the overlap $\Tr{\ket{\mathrm{cat}_\beta}\bra{\mathrm{cat}_\beta}\rho(t)}$
between the quantum state of the $\hat{a}_v$-cavity and the cat state
\begin{align}\label{eq:betaCat}
\ket{\mathrm{cat}_\beta} = \frac{1}{\sqrt{2}} \left(\ket{\beta}+\ket{-\beta}\right),
\end{align}
where the complex amplitude $\beta = 2.0\e{-0.31i\pi}$ is determined by numerical optimization of the final state overlap.
%with $\beta = \sqrt{2.01}\e{-0.03 i \pi}$ which is the amplitude found to match best by Ref.~\cite{PhysRevA.99.023838} for these parameters \footnote{The sign of the phase is, however, opposite in our setup which is maybe a bit puzzling? This should be fixed with the complex conjugation}. Our own numerical optimization yields the same values.  
The fidelity saturates to 
$\Tr{\ket{\mathrm{cat}_\beta}\bra{\mathrm{cat}_\beta}\rho(t)}\simeq 0.98$ and the cat state indeed contains
\begin{align}
n_\beta = |\beta|^2\frac{1-\e{-2|\beta|^2}}{1+\e{-2|\beta^2|}} = |\beta|^2\tanh\left(|\beta|^2\right) = 4.03 
\end{align}
photons.%, which apart from a slight deviation on the third decimal is also what Ref.~\cite{PhysRevA.99.023838} finds. 
The emergence of the travelling cat state is furthermore illustrated by the characteristic Wigner function of the $\hat{a}_v$-cavity mode content, shown at different points in time in \fref{fig:catA}(b). In Ref.~\cite{PhysRevA.99.023838} the performance is further improved by employing a shortcut to adiabaticity which we shall not pursue here.
Notice that with our formalism, we can easily study larger cats without exhausting our computational resources.
%%%%%%%%%%%%%%%%%%%%%%%%%%%%%%%%%%%%%%%%%%%%%%%%%%%%%%%%%%%%%%%%
We emphasize that while the $g^{(1)}(t,t')$-analysis assigns importance to a mode $v_i(t)$ according to its population $n_i$, one can envisage applications with other attributes of interest. For instance, larger Wigner function negativity and stronger quantum correlations and entanglement between subsystems may appear in modes $\tilde{v}(t)$ which do not necessarily hold the largest number of quanta.  Such optimal modes may be identified by optimization of the desired property within subspaces of the complete set eigenmodes of $g^{(1)}(t,t')$.

\section{Generalization to multiple input and output modes}\label{sec:multimode}

\subsection{Cascaded master equation with multiple virtual cavities}
In the appendix, we show how classical wave theory readily provides a model where multiple wave packets can be either emitted or absorbed by cascaded arrays of suitably switched cavities. Our theory employs these cavity modes to solve the corresponding cascaded master equation for the combined system of the input oscillator pulse modes $\{u_i(t)\}_{i=1}^n$, the localized quantum system(s), and the output oscillator pulse modes $\{v_i(t)\}_{i=1}^m$.
The multi-mode extension of our theory thus incorporates $n+m$ virtual cavities, and in the appendix, we describe how the corresponding time dependent coupling strengths $g_{u_i}(t)$ and $g_{v_i}(t)$ are found from classical wave theory applicable to the linear coupling of bosonic fields.
The coupling strengths are evaluated \emph{prior} to solution of the quantum master equation~(\ref{eq:me}) with a Hamiltonian of the form \cite{doi:10.1080/23746149.2017.1343097}
\begin{widetext}
\begin{align}\label{eq:Hfull}
\begin{split}
\H(t) = \H_{\mathrm{s}}(t)
+\frac{i}{2}\left[
\sum_{i=1}^n g_{u_i}^*(t)\ad_{u_i}\left(\sqrt{\gamma}\c
+\sum_{j=1}^{i-1} g_{u_j}(t)\a_{u_j}+\sum_{j=1}^m g_{v_j}(t)\a_{v_j}\right)
+
\sum_{i=1}^m\left(\sqrt{\gamma^*}\Cd + \sum_{j=1}^{i-1} g_{v_j}^*(t)\a_{v_j}^\dagger\right)g_{v_i}(t) \a_{v_i}
-\mathrm{h.c}
\right].
\end{split}
\end{align}
\end{widetext}
and a loss term $\mathcal{D}[\hat{L_0}]\rho$ with Lindblad operator 
\begin{align}\label{eq:Lfull}
\hat{L}_0 = \sqrt{\gamma}\c+\sum_{i=1}^n g_{u_i}(t)\a_{u_i}+\sum_{i=1}^m g_{v_i}(t)\a_{v_i}
\end{align}
along with the damping and decoherence terms $\sum_{i=1}^n \mathcal{D}[\hat{L_i}]\rho$.

\begin{figure*}
\centering
\includegraphics[trim=0 0 0 0,width=1\textwidth]{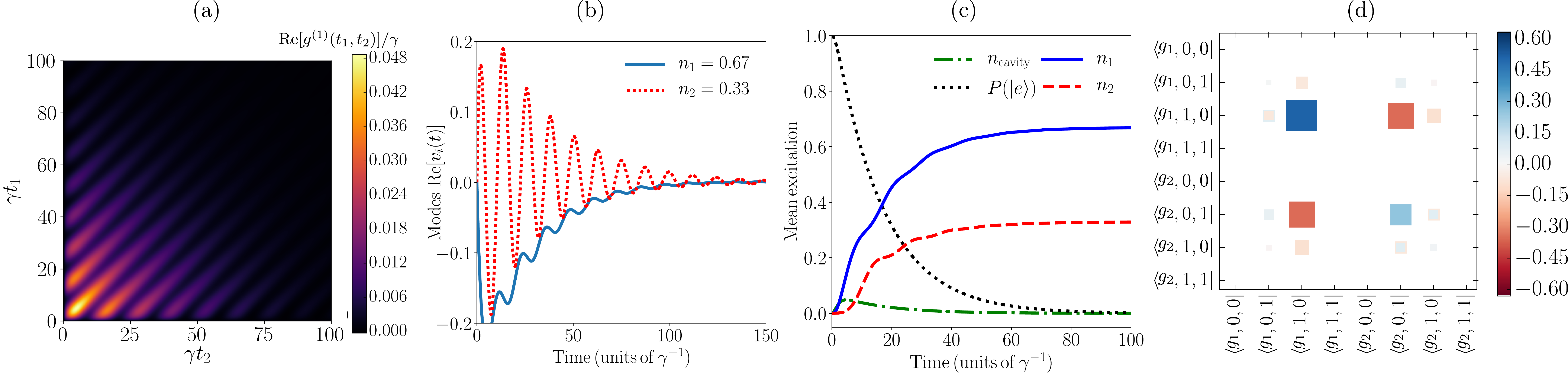}
\caption{
Decaying $\Lambda$ system in a cavity.
(a) 
Real part of the autocorrelation function $g^{(1)}(t_1,t_2)$ for the field emitted by the cavity. 
(b)
The two orthogonal eigenmodes $v_1(t)$ and $v_2(t)$ (real parts shown) of $g^{(1)}(t_1,t_2)$ span the full Hilbert space of the emitted field with respective final mean populations $n_1=0.67$ and $n_2=0.33$.
(c)
Mean excitation in the cavity mode ($n_{\mathrm{cavity}}$), the atom ($P(\ket{e})$ ) and the two output pseudo-cavity detector modes ($n_1$, $n_2$) as functions of time during the two-channel decay process.
(d) Hinton diagramme illustrating the collective state of the atom, the mode $v_1(t)$ and the mode $v_2(t)$
at the end of the decay process (at the time $t = 150 \kappa^{-1}$). 
Results are displayed in a $\ket{\psi_{\mathrm{atom}},\psi_{v_1},\psi_{v_2}}$ eigenstate-basis where the state amplitudes are real.
}
\label{fig:Lambda}
\end{figure*}

\subsection{Photon number and mode entanglement with a quantum emitter}
As an example of a situation with a finite number of relevant output modes, we consider a $\Lambda$-type system with two ground states $\ket{g_1}$ and $\ket{g_2}$ and one excited state $\ket{e}$ in a one-sided cavity with constant outcoupling $\gamma$.
The transitions $\ket{g_1}\leftrightarrow \ket{e} $ and $\ket{g_2}\leftrightarrow \ket{e}$ both couple to the same cavity mode $\a$ with the strength $g = 0.1\gamma$ but the $\ket{g_2}\leftrightarrow \ket{e}$ transition is detuned by $\omega_{12} = 0.5\gamma$ from the cavity resonance. We initialize the $\Lambda$ system in its excited state $\ket{e}$ and observe the decay through the cavity mode.

The results  are displayed in \fref{fig:Lambda}.
The color plot in (a) shows the $g^{(1)}(t_1,t_2)$-autocorrelation function of the field emitted by the cavity. An eigendecomposition of $g^{(1)}(t_1,t_2)$ reveals that only the two modes $v_1(t)$ and $v_2(t)$ in (b) are populated at the final time. Their populations are given by the corresponding eigenvalues, $n_1 = 0.67$ and $n_2 = 0.33$.
Upon identifying these modes, the formalism in Sec.~\ref{sec:multimode} allows us to perform a full quantum simulation of the atomic decay and emission of light into the modes. 

 Figure~\ref{fig:Lambda}(c) shows that while the excitation $P_e$ of the $\Lambda$-system decreases, a small excitation builds up in the cavity field and couples to form the final populations $n_1(t\rightarrow \infty) = 0.67$ and $n_2(t\rightarrow \infty) = 0.33$

While the $\Lambda$-system features two transitions, it is not obvious that they correspond directly to the two eigenmodes of the output field correlation function. During the emission, which lasts about $\gamma/g^2 \simeq 50 \gamma^{-1}$ in our example, however, a frequency difference of $0.5\gamma$ is discernible in the signal, and the two orthogonal eigenmodes are closely associated with emission by the separate atomic transitions. This is confirmed by the correlation between the occupation of the atomic final states and modes shown in the Hinton diagram in \fref{fig:Lambda}(d). 
The final state is approximately
\begin{align}
\ket{\psi(t\rightarrow \infty)}=c_{g_1 1 0}\ket{g_1}\ket{1_{v_1}}\ket{0_{v_2}}+ c_{g_1 0 1}\ket{g_2}\ket{0_{v_1}}\ket{1_{v_2}},
\end{align}
where $|c_{g_1 1 0}|^2 = 0.645$ and $|c_{g_1 0 1}|^2 = 0.317
$. The discrepancy between these numbers and $n_1$ and $n_2$, and the small components in $\ket{g_1}\ket{1_{v_1}}\ket{0_{v_2}}$ and $\ket{g_2}\ket{0_{v_1}}\ket{1_{v_2}}$ seen in the Hinton diagram reflect the small overlap between the actual pulses emitted on the two transitions. For a smaller detuning between the transitions, this overlap becomes larger and a single eigenmode would be predominantly populated and correlated with a superposition of the atomic states in the Hinton diagram.

\section{Photon blockade}
As another application of our theory, we consider the proposal to use an atom in a cavity as a non-linear quantum filter that transmits single photon number states and reflects pulses with higher photon numbers. Many theoretical proposals for such operations exist, and continuous wave experiments have confirmed the anticipated photon anti-bunching  after transmission of a coherent, continuous wave  beam through the proposed devices \cite{birnbaum2005photon, schuster2008nonlinear,RevModPhys.87.1379,hacker2019deterministic}. We are now able to present a theoretical treatment of the modification of quantum pulses of radiation by such filters. 

\subsection{Theoretical model}
We consider a quantum pulse prepared in a state $\ket{\psi_{\mathrm{u}}}$ incident on a symmetric, two-sided cavity resonantly coupled to a qubit system with two states $\ket{g}$ and $\ket{e}$.
In a frame rotating at the cavity frequency, the system is described by the Jaynes-Cummmings Hamiltonian \eqref{eq:HJC}. 
In the experiment of Ref.~\cite{birnbaum2005photon},
a field, detuned by $\omega=g$ from the cavity resonance, is injected into the cavity so that a single incident photon is resonant with the one-excitation dressed state of the system, while, e.g., an $n$-photon state is detuned from the $n$-excitation dressed state by $(n-\sqrt{n})g$. For large $g$, this should lead to off-resonant reflection of the two and higher photon number components, while the one photon component experiences a resonant cavity and is fully transmitted. We study the situation where the incoming field is described by a Gaussian pulse~(\ref{eq:uGauss}) of finite duration $2\tau$, arriving at a time $t_p$, with a frequency modulation factor $e^{-ig t}$ to account for its detuning $g$.  We assume equal transmission rates $\kappa$ of the two cavity mirrors.

A simple extension of our theory is necessary to accommodate the reflection and transmission channels. To represent the  transmitted wave packet $w(t)$ [see schematic in Fig.1(d)], we thus supplement the $\hat{a}_v$ cavity (reflection) with a cavity mode in the transmission channel with annihilation operator $\a_w$ and coupling strength $g_w(t)$ .

In the SLH formalism \cite{doi:10.1080/23746149.2017.1343097}, we find that the combined network evolves according to the Hamiltonian
\begin{widetext}
\begin{align}
\H(t) = \H_{\mathrm{s}}(t) 
+\frac{i}{2}\left[\sqrt{\kappa}\left(g_u^*(t)\a_u^\dagger \c+ g_v(t)\Cd \a_v\right) 
+ \sqrt{\kappa} g_w(t) \Cd \a_w
+ g_u^*(t)g_v(t) \ad_u\a_v
-\mathrm{h.c.}
\right],    
\end{align}
\end{widetext}
and that the damping terms in \eqref{eq:me} must include two Lindblad operators
\begin{align}
\hat{L}_r(t) &= \sqrt{\kappa}\c+g_u(t)\a_u+g_v(t)\a_v,
\\
\hat{L}_t(t) &= \sqrt{\kappa}\c+g_w(t)\a_w.
\end{align}
The former, $\hat{L}_r(t)$ accounts for the part of the radiation from the cavity system interfering with the reflected input signal and appearing in modes orthogonal to the $\hat{a}_v$ detector mode, and $\hat{L}_t(t)$ accounts for the part of the transmitted signal appearing in modes orthogonal to the $\hat{a}_w$ detector mode.
%As an example of great relevance for quantum information distribution, we consider the selective reflection or transmission of a quantum pulse prepared in a state $\ket{\psi_{\mathrm{pulse}}}$ incident on a Fabry-Perot cavity containing a qubit system with two states $\ket{\downarrow}$ and $\ket{\uparrow}$. We assume that $\ket{\downarrow}$ is uncoupled from the cavity mode $\a$, while $\ket{\uparrow}$ is coupled by an interaction $\H_{\mathrm{s}} = ig(\a\ket{e}\bra{\uparrow}-\ad\ket{\uparrow}\bra{e})$ through an ancillary excited state $\ket{e}$.
%In the strong coupling regime, this implies that a qubit in the coupled state $\ket{\uparrow}$ splits the cavity resonance such
%that an incoming pulse resonant at the bare cavity frequency is effectively non-resonantly scattered, yielding perfect reflection with no transmission.
%A qubit in $\ket{\downarrow}$, on the other hand, allows resonant scattering and we expect perfect transmission with a $\pi$ phase shift 

%These arguments assume that the pulse duration $\tau$ is kept well above the cavity lifetime  $~1/(\gamma_r+\gamma_t)$, such that no spectral deformation occurs (see, e.g., Ref.~\cite{PhysRevA.98.030302} for details).

%Figure~\ref{fig:refTrans} shows results for Gaussian input and output pulses $u(t)=v(t)=w(t)$. 
%Parameters are given in the figure caption.

\subsection{Transmission of a single photon}

If the incoming pulse is prepared in a one-photon state, $\ket{\psi_{\mathrm{u}}} = \ket{1}$, the linearity of the resulting equations of motion in the single excitation subspace allows exact solution of the scattering problem \cite{PhysRevA.98.030302}. The frequency dependent transmission coefficient is
\begin{align}\label{eq:T}
T(\omega) = \frac{i\omega}{(g^2-\omega^2)-i\kappa\omega}
\end{align}
and the reflection coefficient $R(\omega) = 1+T(\omega)$, such that after the scattering process, the incoming one-photon pulse is split into a  transmitted mode $w(t)$ with population $\braket{\ad_w \a_w} = \int d\omega\,|T(\omega)u(\omega)|^2$ and a reflected mode $v(t)$ with population $\braket{\ad_v \a_v} = \int d\omega\,|R(\omega)u(\omega)|^2$, where $u(\omega)$ is the Fourier transform of the incoming mode~(\ref{eq:uGauss}), multiplied by the time-dependent phase factor $\e{-igt}$ to represent a carrier detuning of $g$.

\begin{figure}[h!]
\centering
\includegraphics[trim=0 0 0 0,width=0.75\columnwidth]{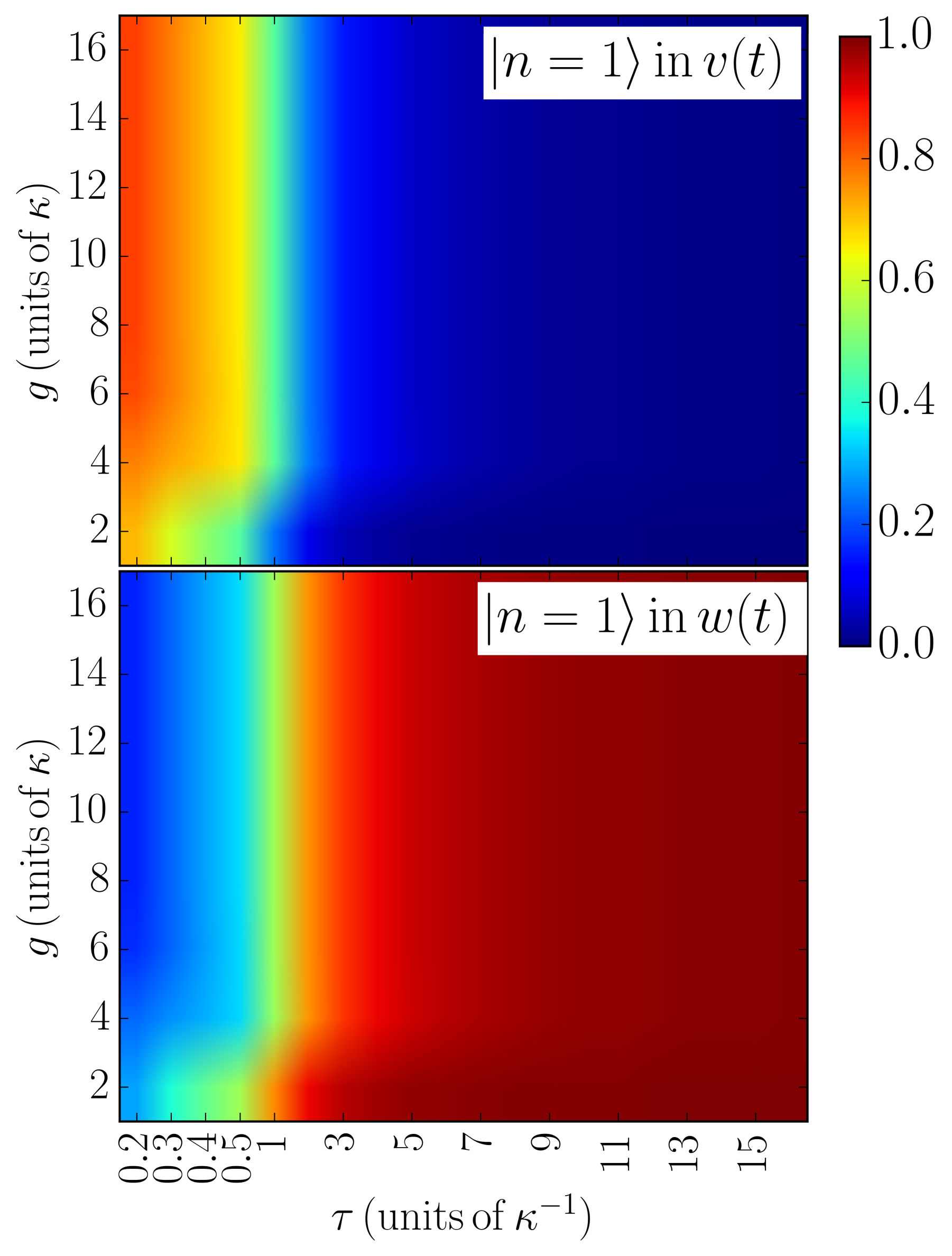}
\caption{
Reflection and transmission of an incoming one-photon state $\ket{\psi_{\mathrm{u}}} = \ket{1}$, with a carrier frequency resonant with the first excited Jaynes-Cummings eigenstate. 
A state $\ket{1}$ populates a superposition of the reflected and transmitted wave packets, $v(t)$ and $w(t)$, depending on the value of $\tau$ and $g$. 
Notice that in order to explore the $\kappa\tau<1$ regime, the $\tau$ scale in the figure is not linear.
}
\label{fig:kimble1}
\end{figure}

Figure~\ref{fig:kimble1} shows these populations for different values of the pulse duration $\tau$ and the coupling $g$. The resonance condition is only valid for the one-photon component when the incident pulse carrier frequency  is tuned exactly $g$ away from the cavity resonance. 
For $\tau< 1/\kappa$, however, the pulse is spectrally broader than the cavity linewidth and frequency components outside $\sim \kappa$ are reflected. At very small $\tau$, this effect is dominating and we see a nearly complete reflection. For $\tau> 1/\kappa$, on the other hand, the incident pulse is spectrally narrow and the desired resonant transmission occurs. The transition between the long and short pulse regimes depends on the value of $g$, as the half width of the transmitted intensity [c.f. \eqref{eq:T}] changes from $\kappa$ for small $g$ to $\kappa/2$ for large $g$.

\subsection{Transmission of higher photon number states}
\begin{figure}[]
\centering
\includegraphics[trim=0 0 0 0,width=0.75\columnwidth]{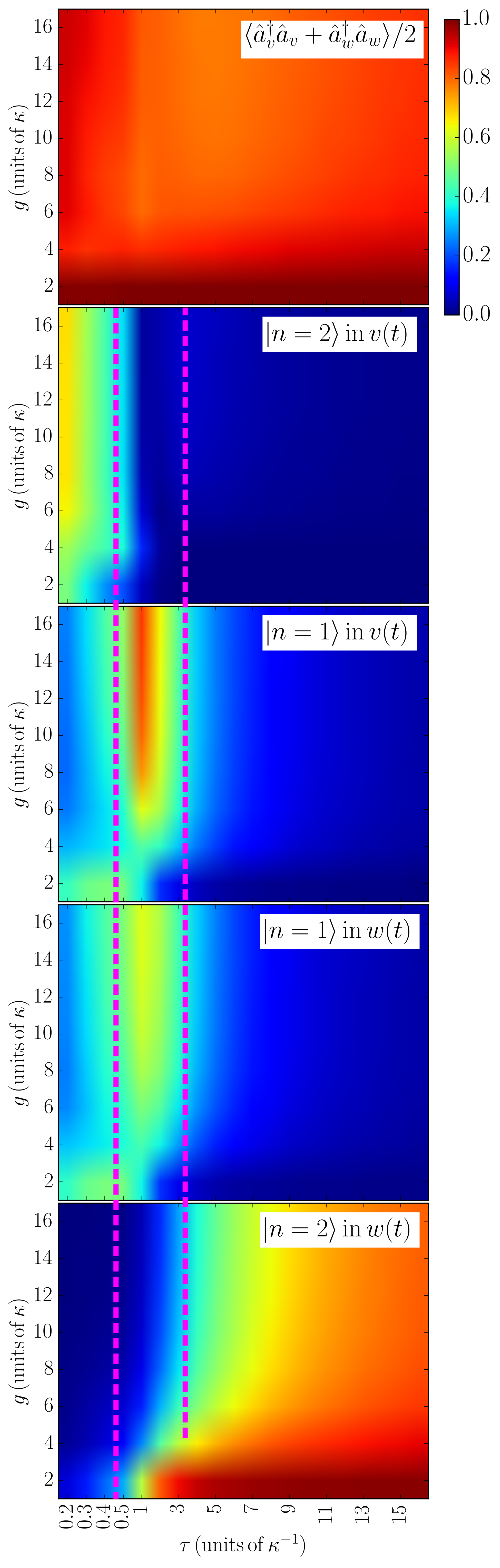}
\caption{
Photon blockaded transmission for an incoming two-photon state $\ket{\psi_{\mathrm{u}}} = \ket{2}$. Upper panel: Relative excitation $\braket{\ad_v\a_v+\ad_w\a_w}/2$ in the most populated reflected $v(t)$ and transmitted $w(t)$ modes as a function of $\tau$ and $g$. 
Four lower panels: One ($\ket{1}$) and two ($\ket{2}$) photon final state populations in the reflected and transmitted pulses $v(t)$ and $w(t)$ for different values of $\tau$ and $g$. Dashed, magenta lines mark different duration intervals discussed in the main text. In the $\kappa\tau<1$ regime, the $\tau$ scale in the figures is not linear .
}
\label{fig:kimble}
\end{figure}

Beyond the one-photon subspace, the problem requires numerical solution which we shall now perform to study the scattering of an incoming two-photon pulse $\ket{\psi_{\mathrm{u}}} = \ket{2}$.
For each value of $\tau$ and $g$, we first determine the output correlation function  as in Sec.~\ref{sec:g1} to identify the two dominant output modes $v(t)$ and $w(t)$ in the reflection and transmission channels, respectively. While for $g=0$, the scattering is linear and occurs into a superposition of a reflected and a transmitted wave packet mode, the non-linearity for $g>0$ causes scattering into additional, orthogonal modes as signified by the decrease in $\braket{\ad_v\a_v+\ad_w\a_w}/2$ shown by the color plot in the upper panel of \fref{fig:kimble}. The retained excitation varies between $77\%$  and $100\%$ as a function of of $g$ and $\tau$.   

The four lower panels of \fref{fig:kimble} show the occupation of the Fock states in the reflected and transmitted pulses for different values of $\tau$ and $g$. For $g\simeq 0$, the linear scattering yields a transformation in the basis of the incoming, the reflected and the transmitted wave packet modes $\ket{\psi_u,\psi_v,\psi_w}$,
\begin{align}\label{eq:BS}
\begin{split}
(\ad_u)^2&\ket{0,0,0} \rightarrow (c_v \ad_v + c_w \ad_w)^2\ket{0,0,0} 
\\
&= \sqrt{2}c_v^2\ket{0,2,0}+ \sqrt{2}c_w^2\ket{0,0,2} + 2 c_v c_w \ket{0,1,1},
\end{split}
\end{align}
where the coefficients $c_v$ and $c_w$ depend on the value of $\kappa\tau$.
For a short incident pulse $\tau< 1/\kappa$, the bulk of the pulse frequency contents is beyond the cavity line width and both photons are reflected ($c_v\simeq 1, c_w\simeq 0$). As $\tau$ approaches $1/\kappa$, the excitation is distributed on the output channels with always equal population of the $\ket{1}$ output components, cf., \eqref{eq:BS}.

For large $g$, we observe three different regimes: For small $\tau< 1/\kappa$, the incoming $\ket{2}$ state is fully reflected. For $\tau\simeq 1/\kappa$, the linear beam splitter relation~(\ref{eq:BS}) no longer applies and the  $\ket{1}$ reflected and transmitted components have different probabilities. 
For $\tau >1/\kappa$, the two-photon pulse is predominately transmitted.

This may seem contrary to the aim of the proposal and the anti-bunching results reported in Ref.~\cite{Birnbaum2005}, which suggest that the $\ket{2}$ state is reflected by the off-resonant qubit-dressed cavity. Our result, however, illustrates the difference between interactions with stationary and travelling photons. The expected dependence on the photon number stems from the non-equidistant spectrum of the Jaynes-Cummings Hamiltonian, and thus assumes that all excitations are simultaneously present in the cavity. 
However, for large $\tau$ the two-photon state is dominated by field components where the excitations are separated in time by more than $1/\kappa$ and may thus pass the cavity sequentially. Only simultaneous presence within the photon lifetime inside the cavity is suppressed and this explains the observed antibunching in continuous wave experiments. For small $\tau \lesssim \kappa^{-1}$, we return to the problem of a pulse that is spectrally broader than the cavity linewidth which therefore reflects the photon irrespective of the photon number.

This example emphasizes a fundamental time-bandwidth dilemma that may easily be overlooked in intuitive arguments for the manipulation of quantum states of light. With the theory presented here, we no longer have recourse to intuitive analogies with Jaynes-Cummings dynamics, and we may develop better and more precise insights in the non-linear dynamics of pulses of quantum radiation. 

\section{A \textit{snow ball} in a thermal channel}
\begin{figure}
\centering
\includegraphics[trim=0 0 0 0,width=1\linewidth]{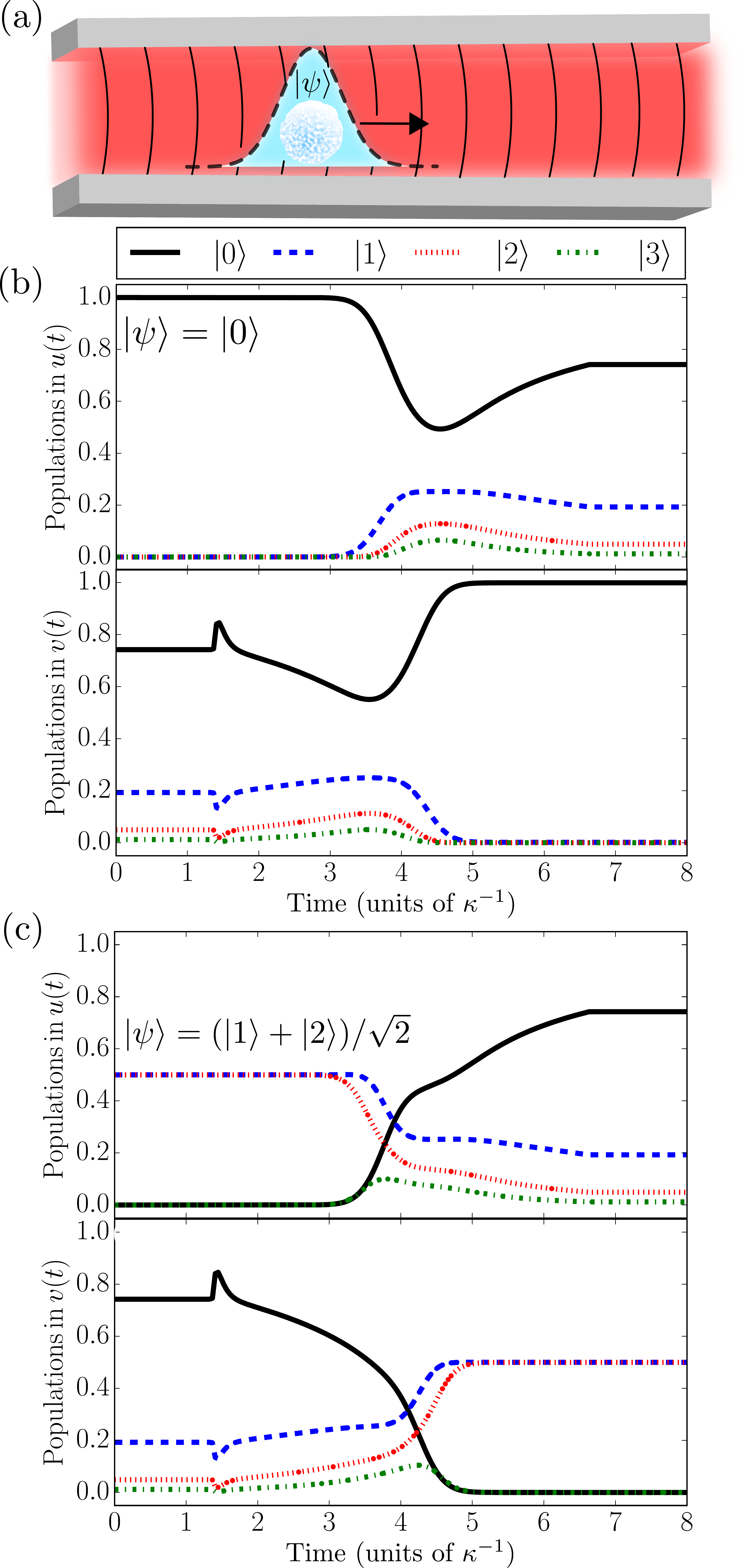}
\caption{Pure state $\ket{\psi}$ pulse propagation in a thermal channel, illustrated in (a).
A Gaussian mode $u(t)$ of width $\tau=0.5\kappa^{-1}$ is transferred through a channel with an incoherent flux of $3\kappa$ thermal photons to be collected by a mode $v(t)$ of the same shape.
The input mode is prepared in a vacuum state $\ket{\psi}=\ket{0}$ in (b) and in a superposition state $\ket{\psi}=(\ket{0}+\ket{1})/\sqrt{2}$ in (c).
Populations in the four lowest Fock states as functions of time are displayed for $u(t)$ in the upper panels and for $v(t)$ in the lower panels.
}
\label{fig:snowball}
\end{figure}
Our theoretical description has so far assumed that the quantum input and output pulses propagate and the local scatterer is situated in a vacuum environment. 
In realistic settings, however, the waveguide and the discrete components may be kept at a finite temperature $T>0$. For microwaves and acoustic waves, the radiation frequency may be so low, that we cannot ignore thermal quanta, and it is interesting to study the transmission of quantum states on the background of such thermal quanta \cite{PhysRevX.7.011035,PhysRevLett.118.133601}, see Fig.~\ref{fig:snowball}(a) for an illustration. We can model a thermal flux of photons by cascading yet another virtual cavity \textit{before} the other components of the system. 
The internal mode $\a_{\mathrm{in}}$ of this cavity is coupled to a thermal environment with a rate $\kappa^\prime$ and to an output line with a rate $\kappa$ to form an equilibrium thermal (exponential) distribution of photons (or phonons). The output field thus mimics a thermal state over a frequency range $\simeq \kappa+\kappa^\prime$, which is supposed to cover the spectrum of the other systems and cavities. The cascaded quantum system composed of the thermal source, an input pulse cavity, a scattering quantum system and a final output pulse cavity obey the master equation \eqref{eq:me} with the additional Hamiltonian,
\begin{align}
\H_{\mathrm{th}} = \frac{i}{2}\left[\sqrt{\kappa}g_u(t)\ad_{\mathrm{in}}\a_u+\sqrt{\kappa\gamma}\ad_{\mathrm{in}}\c+\sqrt{\kappa}g_v(t)\ad_{\mathrm{in}}\a_v-\mathrm{h.c.} \right]    
\end{align}
and Lindblad operators  
\begin{align}
\hat{L}_0(t) &= \sqrt{\gamma}\c+g_u(t)\a_u+g_v(t)\a_v+ \sqrt{\kappa}\a_{\mathrm{in}},
\\
\hat{L}_{+} &= \sqrt{\tilde{N}\kappa'}\ad_{\mathrm{in}}
\\
\hat{L}_{-} &= \sqrt{(\tilde{N}+1)\kappa'}\a_{\mathrm{in}}.
\end{align}
The mean photon number in the input thermal cavity is $N = \tilde{N}/(1+\kappa/\kappa^\prime)$, and the flux of thermal photons incident on the subsequent systems is $\kappa N$.

It was recently shown theoretically that it is possible to transmit quantum states between cavities through a thermally excited channel \cite{PhysRevX.7.011035,PhysRevLett.118.133601}. We shall now apply our formalism to investigate the same setup, and we note that we may readily proceed to other physical systems, such as qubits and non-linear devices. We now deal with modelling of real cavities coupled to the waveguide with the aim to transfer a quantum state $\ket{\psi}$. Perfect transmission is ensured if the coupling coefficients $g_u(t)$ and $g_v(t)$, match the same travelling pulse shape $u(t)=v(t)$, and while there may be thermal quanta propagating alongside the pulse, they occupy orthogonal modes and are hence not captured by the receiving cavity. To be precise, all spectrally relevant modes are thermally excited in the waveguide, but the coupling that releases a quantum state from the $\hat{a}_u$ cavity into a travelling wave packet causes the initial thermal content of that same wave packet in the waveguide to enter and occupy the $\hat{a}_u$ cavity, i.e., the initial quantum state of the cavity and the thermal state of the pulse are swapped.

In Fig.~\ref{fig:snowball}, we show the results for a Gaussian mode~(\ref{eq:uGauss}) of width $\tau = 0.5\kappa^{-1}$ propagating in a waveguide illuminated by an incoherent photon flux of  $N\kappa = 3\kappa$ from a thermally excited cavity. Panel (b) shows what we may colloquially call a \textit{snowball in Hell}: a pulse prepared in the vacuum state $\ket{\psi_u} = \ket{0}$ is sent through a much warmer channel. As this state replaces the initial thermal state of the $\hat{a}_v$ cavity, it effectively cools that system. Figure~\ref{fig:snowball}(b) shows that the state of $v(t)$ is replaced by the vacuum state at the final time, and hence, the travelling wave packet (the \textit{snowball}) is not heated during the propagation. Fig.~\ref{fig:snowball}(c) illustrates how a pulse prepared in a superposition state $\ket{\psi_u} = (\ket{1}+\ket{2})/\sqrt{2}$ may similarly be transferred through a thermal channel without loss of fidelity. The transient spikes in the populations in the $\hat{a}_v$ cavity are artefacts due the abrupt and hence broad bandwidth coupling to vacuum frequency components outside the finite ($\kappa + \kappa'$) bandwidth of our ``thermal'' bath.  

We imagine that transmission of {\it snow ball} vacuum states, prepared, e.g., in a heralded manner \cite{PhysRevLett.109.050507}, may be employed to cool a finite number of quantum degrees of freedom in a more economical manner than present days' cooling of entire bulk system. Thus, a finite number of super conducting oscillator and qubit degrees of freedom used in quantum computing may be kept at few m$K$ temperatures by a supply of \textit{snow balls}, while the surrounding apparatus may be kept at few $K$ to ensure superconductivity.

\section{Discussion}
In this manuscript, we have developed a quantum theory of pulses of radiation that can be adopted and generalized to accommodate a number of scenarios in quantum optics and quantum information applications. Our examples emphasize important differences between stationary modes and travelling pulses and raise awareness against too direct application of single and few mode formalism and intuition for the propagation and manipulation of travelling states. The theory presents numerous options to discard the emitted field components, determine their mean properties, or calculate their full quantum state by a cascaded density matrix theory. We may thus recover and extend established theories as well as address new problems within one and the same theoretical framework. 

While examples with light and microwave pulses may come first to mind, the theory applies equally to acoustic waves coupled, e.g., by piezoelectric interactions to circuit QED components \cite{Bienfait368,Ekstr_m_2019} and we also imagine applications with other wave phenomena such as Bogoliubov excitations in cold gases, spin waves, etc. 

We have far from exhausted the theoretical possibilities of the formalism, and we expect progress in a number of directions. For instance, pulses propagating in an extended, non-linear medium may be investigated by representing the suitably discretized medium by a matrix product state \cite{C4FD00206G,manzoni2017simulating,mahmoodian2019dynamics}, cascaded between the input and output pulse cavity modes. We may thus perform accurate calculations of non-linear, photon number dependent dispersion effects and, e.g., pursue splitting of incoming pulses according to their Fock state components.

Another natural direction of research is dynamics where non-classical pulses interact with a quantum system and are subsequently detected
\cite{PhysRevLett.70.2273,PhysRevA.86.043819,PhysRevA.96.023819,fischer2018scattering,Gough_2014}. Advanced detection schemes, employing incident squeezed, Fock or Schr\"odinger cat states for precision metrology, may thus be treated in an exact manner.

\begin{acknowledgments}
The authors would like to thank David Petrosyan for helpful comments on the manuscript and acknowledge support from the European Union FETFLAG program, Grant No. 820391 (SQUARE), and the U.S. ARL-CDQI program through cooperative Agreement No. W911NF-15-2-0061.
\end{acknowledgments}

%%%%%%%%%%%%%%%%%%%%%%%%%%%%%%%%%%%%%%%%%%%%%%%%%%%%

\section*{appendix}\label{sec:Appendix}
\appendix

In Sec.~\ref{sec:multimode}, we explain how our formalism may be extended to accommodate several output and input modes. Here we describe in detail how this is accomplished and derive the time dependent coupling strengths $g_{u_i}(t)$ and $g_{v_i}(t)$, appearing in Eqs.~(\ref{eq:Hfull})~and~(\ref{eq:Lfull}).

\begin{figure}[h!]
\centering
\hspace{2.5em}
\includegraphics[trim=0 0 0 0,width=1.0\columnwidth]{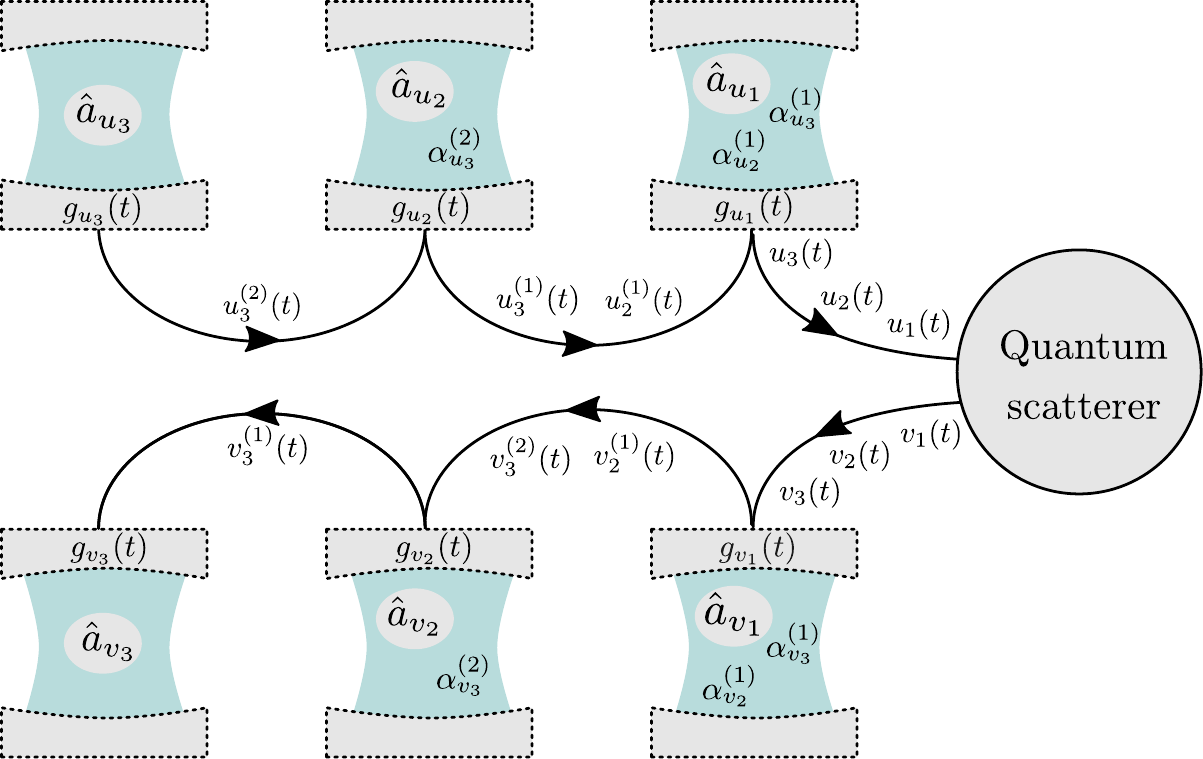}
\caption{Illustration of the extension of our formalism to include (for instance) $n=3$ incoming modes, $u_1(t), u_2(t)$ and $u_3(t)$, and $m=3$ outgoing modes $v_1(t)$, $v_2(t)$ and $v_3(t)$.
The quantum state content of incoming (outgoing) mode $i$ is represented by a mode $\hat{a}_{u(v)_i}$ in a virtual cavity with coupling $g_{u(v)_i}(t)$.
During the reflections, the modes are reshaped, and, e.g., input cavity three must emit the mode $u_3^{(2)}(t)$, which is transformed into  $u_3^{(1)}(t)$ and subsequently  into the desired $u_3(t)$ pulse incident on the scatterer (see main text for details).
}
\label{fig:multiMode}
\end{figure}

To extend our formalism to include $m$ orthogonal output modes $\{v_i(t)\}_{i=1}^m$, we assume that after the first virtual cavity, which perfectly absorbs the mode $v_1(t)$, the field is serially reflected on a sequence of virtual cavities. They each have their own coupling strength $g_{v_{i}}(t)$, designed such that the quantum state content of the mode $v_i(t)$ is precisely captured by the internal field $\a_{v_i}$. 
This idea is illustrated in \fref{fig:multiMode} for three output modes.

To model the scattering of $n$ orthogonal incoming modes $\{u_i(t)\}_{i=1}^{n}$, we consider likewise a sequence of cascaded virtual input cavities with coupling strengths $g_{u_i}(t)$ and internal fields $\a_{u_i}$. The final of these directly ejects the first mode $u_1(t)$ towards the scatterer, while previous ones eject modes which are serially reflected on every cavity until the scatterer is reached.
This is illustrated in \fref{fig:multiMode} for the case of three input modes.

To determine the correct time dependent cavity coupling strengths $g_{v_i}(t)$ and $g_{u_i}(t)$ of the $n+m$ virtual cavities corresponding to the desired modes, we must take into account the distortion of each pulse shape by reflections on the subsequent sequence of cavities. Due to the linearity of the virtual cavity systems, this can be accomplished by the evolution of classical mode amplitudes.
We present the detailed derivation of the cavity couplings in the following, and we emphasize that this calculation is performed prior to and independent of the solution of the ensuing cascaded quantum master equation.

\subsection{Coupling strengths for multiple outputs}
In the scheme outlined above and illustrated in \fref{fig:multiMode},
we must take into account that the output modes are reshaped by each reflection. That is, after the $j$th (virtual) cavity, the remaining modes are transformed as $v_i(t) \rightarrow v^{(j)}_i(t)$, where, since the reflection is a unitary process, the orthogonality between modes is preserved.

Let us considering the output mode $v_2(t)$.
During the reflection of this mode on the first virtual cavity, the contribution $\alpha_{v_2}^{(1)}$ to the cavity field amplitude due to this particular pulse solves a differential equation
\begin{align}\label{eq:alphav21}
\dot{\alpha}_{v_2}^{(1)} = -g_{v_1} v_2 -\frac{|g_{v_1}|^2}{2}\alpha_{v_2}^{(1)}
\end{align}
from an initial value $\alpha_{v_2}^{(1)}(0) = 0$.
In \eqref{eq:alphav21} and below, we omit the explicit time dependence of the rates and modes for simplicity of notation.
The corresponding reflected mode amplitude is given by
\begin{align}
v_2^{(1)} = v_2+g_{v_1}^*\alpha_{v_2}^{(1)}.
\end{align}
In order to associate the internal mode $\a_{v_2}$ of the second cavity with the mode $v_2(t)$, scattered from the quantum system, we should hence define the coupling rate of the second virtual cavity as [Eq.~(3) of the main text]
\begin{align}
g_{v_2}(t) = -\frac{\left[v^{(1)}_{2}(t)\right]^*}{\sqrt{\int_0^t d\tp\, |v^{(1)}_2(\tp)|^2}}.
\end{align}

Likewise, a third mode $v_3(t)$ contributes an amplitude in the first cavity given by 
\begin{align}
\dot{\alpha}_{v_3}^{(1)} = -g_{v_1} v_3 -\frac{|g_{v_1}|^2}{2}\alpha_{v_3}^{(1)},
\end{align}
and is rescattered to the second cavity as
\begin{align}\label{eq:v31}
v_3^{(1)} = v_3+g_{v_1}^*\alpha_{v_3}^{(1)}.
\end{align}
In the second cavity, a corresponding amplitude $\alpha_{v_3}^{(2)}(t)$ then builds up according to 
\begin{align}
\dot{\alpha}_{v_3}^{(2)} &= -g_{v_2} v_3^{(1)} -\frac{|g_{v_2}|^2}{2}\alpha_{v_3}^{(2)}  \nonumber
\\ 
&= -g_{v_2} \left(v_3+g_{v_1}^*\alpha_{v_3}^{(1)}\right) -\frac{|g_{v_2}|^2}{2}\alpha_{v_3}^{(2)},
\end{align}
where we applied \eqref{eq:v31} in the final equation.
The reshaped mode, arriving at the third cavity, is 
$v_3^{(2)} = v_3^{(1)}+g_{v_2}^*\alpha_{v_3}^{(2)} = v_3+g_{v_1}^*\alpha_{v_3}^{(1)}+g_{v_2}^*\alpha_{v_3}^{(2)}$, which defines the coupling strength 
$g_{v_3}(t) = -[v^{(2)}_{3}(t)]^*/\sqrt{\int_0^t d\tp\, |v^{(2)}_3(\tp)|^2}$ to the associated cavity mode $\a_{v_3}$.

By now, the generalization to $m$ modes should be clear. 
For mode $1<i\leq m$, we should solve
$i-1$ coupled differential equations
\begin{align}\label{eq:alphaEqs}
\dot{\alpha}_{v_i}^{(j)} = -g_{v_j} \left(v_i+\sum_{k = 1}^{j-1} g_{v_k}^*\alpha_{v_i}^{(k)}\right) -\frac{|g_{v_j}|^2}{2}\alpha_{v_i}^{(j)}
\end{align}
for the associated amplitudes $\alpha_{v_i}^{(j)}(t)$ with $j=1,2,...,i-1$ in each virtual cavity prior to the $i$th cavity.

Then, the mode in that cavity $\a_{v_i}$ captures precisely the quantum state of the original mode $v_i(t)$ if 
\begin{align}
g_{v_i}(t) = -\frac{\left[v^{(i-1)}_{i}(t)\right]^*}{\sqrt{\int_0^t d\tp\, |v^{(i-1)}_i(\tp)|^2}}
\end{align}
with
\begin{align}
v^{(i-1)}_{i} = v_i+\sum_{k = 1}^{i-1} g_{v_k}^*\alpha_{v_i}^{(k)}.
\end{align}

We note that with $m$ modes, one needs to solve $\sum_{i=1}^m (i-1) = m(m-1)/2$ differential equations for the needed amplitudes $\alpha_{v_i}^{(j)}(t)$.
Due to the increased Hilbert space dimension of the density matrix, we do not imagine that the present formalism will find applications for more than a few input and output modes.

\subsection{Coupling strengths for multiple inputs}
As illustrated in \fref{fig:multiMode}, the input modes similarly experience reflections which cause unitary transformations before they reach their final destination at the scatterer.
By $u_i^{(j)}(t)$ we denote the shape of the mode $u_i(t)$ just \textit{before} it is reflected on cavity $j$ (counting the cavities from the scatterer and out).
In order to associate the field $\a_{u_i}$ in each cavity with a mode $u_i(t)$ \textit{arriving} at the scatterer, 
the coupling strength $g_{u_i}(t)$ must thus be designed such that the mode $u_i^{(i-1)}(t)$, actually ejected from the $i$th towards the $(i-1)$th virtual cavity, correctly transforms into $u_i(t)$. That is [Eq.~(2) of the main text]
\begin{align}\label{eq:gui}
g_{u_i}(t) = \frac{\left[u^{(i-1)}_{i}(t)\right]^*}{\sqrt{1-\int_0^t d\tp\, |u^{(i-1)}_i(\tp)|^2}}.
\end{align}

The $u_i^{(i-1)}(t)$ are determined by propagating backwards from the scatterer.
For instance, a second mode $u_2(t)$ is ejected from the second virtual cavity as $u_2^{(1)}(t)$ and during reflection on the first cavity, the cavity amplitude, $\alpha_{u_2}^{(1)}$ solves the equation
\begin{align}
\dot{\alpha}_{u_2}^{(1)} &= -g_{u_1} u_2^{(1)} -\frac{|g_{u_1}|^2}{2}\alpha_{u_2}^{(1)}.
\end{align}
The reflected mode is required to produce the desired shape,  $u_2 = u_2^{(1)}+g_{u_1}^*\alpha_{u_2}^{(1)}$. 
The amplitude equation may thus be rewritten
\begin{align}
\dot{\alpha}_{u_2}^{(1)} &= -g_{u_1} u_2 +\frac{|g_{u_1}|^2}{2}\alpha_{u_2}^{(1)}
\end{align}
and solved. The emitted pulse is given by $u_2^{(1)} = u_2-g_{u_1}^*\alpha_{u_2}^{(1)}$ and the coupling strength follows from ~(\ref{eq:gui}).

For a third input mode $u_3(t)$, the corresponding mode $u_3^{(2)}(t)$, ejected from the third virtual cavity, is reflected on the second and first virtual cavitites before reaching the scatterer. During these reflections, amplitude contributions $\alpha_{u_3}^{(2)}$ and $\alpha_{u_3}^{(1)}$ build up inside those cavities according to the equations
\begin{align}\label{eq:alpha_u3}
\begin{split}
\dot{\alpha}_{u_3}^{(1)} &= -g_{u_1} u_3^{(1)} -\frac{|g_{u_1}|^2}{2}\alpha_{u_3}^{(1)}
\\
\dot{\alpha}_{u_3}^{(2)} &= -g_{u_2} u_3^{(2)} -\frac{|g_{u_2}|^2}{2}\alpha_{u_3}^{(2)}.
\end{split}
\end{align}
The output from the first virtual cavity is given by $u_3^{(1)}+g_{u_1}^*\alpha_{u_3}^{(1)}$, and
we require this to yield the desired mode, $
u_3 = u_3^{(1)}+g_{u_1}^*\alpha_{u_3}^{(1)}.
$
At the same time, the input to the first cavity from the second cavity is given by
$
u_3^{(1)}(t) = u_3^{(2)}+g_{u_2}^*\alpha_{u_3}^{(2)} 
$. 
These relations allow us to rewrite Eqs.~(\ref{eq:alpha_u3}) in terms of the mode $u_3(t)$:
\begin{align}\label{eq:alpha_u3_two}
\begin{split}
\dot{\alpha}_{u_3}^{(1)} &= -g_{u_1} u_3 +\frac{|g_{u_1}|^2}{2}\alpha_{u_3}^{(1)}
\\
\dot{\alpha}_{u_3}^{(2)} &= -g_{u_2} (u_3-g^*_{u_1}\alpha_{u_3}^{(1)}) +\frac{|g_{u_2}|^2}{2}\alpha_{u_3}^{(2)}.
\end{split}
\end{align}
Upon solving the coupled differential equations~(\ref{eq:alpha_u3_two}), we obtain the emitted pulse, $u_3^{(2)} = u_3-g^*_{u_1}\alpha_{u_3}^{(1)}-g^*_{u_2}\alpha_{u_3}^{(2)}$, and the coupling strength of the third cavity, $g_{u_3}(t)$ follows from \eqref{eq:gui}. 

Extending this line of thought reveals that 
$n$ input modes may be incorporated by solving $n(n-1)/2$ differential equations 
\begin{align}
\dot{\alpha}_{u_i}^{(j)} = -g_{u_j} \left(u_i-\sum_{k=1}^{j-1}g^*_{u_i}\alpha_{u_i^{(k)}}\right) -\frac{|g_{u_i}|^2}{2}\alpha_{u_i}^{(j)},
\end{align}
yielding the temporal mode shapes 
\begin{align}
u_i^{(i-1)} = u_i - \sum_{k=1}^{i-1} g^*_{u_k}\alpha_{u_i}^{(k)}
\end{align}
which define the coupling strengths in \eqref{eq:gui}.

%\bibliography{master}{}
%merlin.mbs apsrev4-1.bst 2010-07-25 4.21a (PWD, AO, DPC) hacked
%Control: key (0)
%Control: author (0) dotless jnrlst
%Control: editor formatted (1) identically to author
%Control: production of article title (0) allowed
%Control: page (1) range
%Control: year (0) verbatim
%Control: production of eprint (0) enabled
%

\end{document}